# Status Quo Bias in Users' Information Systems (IS) Adoption and Continuance Intentions: A Literature Review and Framework

**Full research paper**

**Saliya Nugawela**
Faculty of Business and Law
Swinburne University of Technology
Melbourne, Australia
Email: snugawela@swin.edu.au

**Darshana Sedera**
Faculty of Business, Law, and Arts
Southern Cross University
Gold Coast, Australia
Email: darshana.sedera@scu.edu.au

## Abstract

Information systems (IS) adoption and continuance intentions of users have a dominant effect on digital transformation in organisations. However, organisations undergoing digital transformation face substantial barriers due to user resistance to IS implementations. Status quo bias (SQB) plays a vital role in users' decision-making regarding adopting new IS or continuing to use existing IS. Despite recent research to validate the effects of SQB on user resistance to IS implementations, how SQB affects the IS adoption and continuance intentions of users remains poorly understood, making it harder to develop ways of successfully dealing with it. To address the gap, we performed a systematic literature review on SQB in IS research. Our proposed framework incorporates the psychological phenomena promoting the status quo, SQB theory constructs, levels of SQB influence, and factors reducing the user resistance to IS implementations to enhance the understanding of IS adoption and continuance intentions.

# 1  Introduction

A clear understanding of the information systems (IS) non-use continuum, which commences from resistance, under-use, non-use, and discontinuance, is essential when technologies and systems change rapidly (Baumer et al. 2014; Soliman and Rinta-Kahila 2020). Introducing an IS into an organisation yields substantial changes to one's work practices, resulting in tensions among employees' work environments (Bala and Venkatesh 2016; Bala and Venkatesh 2013; Sedera and Lokuge 2017). Such tensions then will have a negative impact on one's job performance and satisfaction (Sedera 2006; Zhang and Venkatesh 2017). Therefore, if the users' IS adoption and continuance intentions are not well-understood, then resistance and disruption behaviours may ensue, causing reduced operational efficacy, decreased motivation, decreased job performance, violation of work ethics, inventing workarounds, resentment, underutilisation, and even sabotaging the system (Recker 2014; Recker 2016). Therein, the Status Quo Bias Theory (SQBT) perspective is widely used to understand information systems users' resistance behaviour and adoption issues of new information systems (Lee and Joshi 2017). Samuelson and Zeckhauser (1988) view the status quo bias as an obstacle in the transition to better alternatives. "*The status quo choice acts as a psychological anchor. Roughly speaking, the stronger the individual's previous commitment to the status quo, the stronger the anchoring effect*" (Samuelson and Zeckhauser 1988, p. 41). "*The phenomenon of status quo bias among IT-savvy employees and its influence on their IT adoption or resistance as a dual phenomenon is underexamined despite their crucial role in technology renewal and IT project success initiatives*" (Shirish and Batuekueno 2021, p. 876). Even though there is great optimism with respect to emerging technology, Hofmann (2020, p. 252), argues that "*there is a status quo bias fuelling irrational attitudes to emergent science and technology and greatly hampering useful development and implementation.*"

This paper aims to better understand current research about the effects of Status quo bias on users' information systems adoption and continuance intentions. Hence, we pose the following research questions:

RQ1: What is the current body of knowledge of SQB in the context of IS Adoption and Continuance?

RQ2: What are the future directions for studying the SQB in IS Adoption and Continuance?

To answer our research questions, we conducted a systematic literature review. We then propose a conceptual framework for studying SQB in IS adoption and continuance based on our results. Our contribution to IS research is twofold. Firstly, our overview and framework can be seen as a step towards developing a better understanding of SQB in IS adoption and continuance. Secondly, our SQB in IS adoption and continuance framework assist in identifying current gaps that can support academics with future research. Further, our understanding of SQB and how the concept is used in the context of IS adoption and continuance may help develop insights into the particularities of successful IS implementation in organisations.

This paper proceeds in the following manner. In the next section, the conceptual background of this research is presented. Then, the study design provides details of the systematic literature review utilised in the study. Next, the results section is provided, followed by the conclusion section. Finally, the practical and research implications are provided in conclusion, along with limitations and future research areas.

# 2  Conceptual Background – Status Quo Bias Theory

"*Status quo bias theory aims to explain people's preference for maintaining their current status or situation*" (Kim and Kankanhalli 2009, p. 569). The study conducted by Samuelson and Zeckhauser (1988) (as cited in Polites and Karahanna 2012) explains why the status quo bias perspective results in "*individuals disproportionately make decisions to continue an incumbent course of action, rather than switching to a new (potentially superior) course of action*" (p. 23). Under the influence of status quo bias, in deciding to follow a new path of action, individual decision-makers are biased toward maintaining the status quo (doing nothing or maintaining one's current or previous decision) (Polites and Karahanna 2012; Samuelson and Zeckhauser 1988). In decision-making practice, status quo bias stems partly from a mental illusion and partly from psychological inclination (Samuelson and Zeckhauser 1988). The individual may prefer the status quo due to convenience, habit, inertia, policy, custom, innate conservatism, attachment to past choice and the fact that he or she may have lived with the status quo choice for some time (Samuelson and Zeckhauser 1988). Samuelson and Zeckhauser (1988) state that SQBT provides eight constructs that lead to SQB. Table 1 describes the eight constructs of the Status Quo Bias Theory.

| SQBT Construct | Definition |
|---|---|
| Loss Aversion | Individuals weigh potential losses heavier than potential gains in making decisions (Samuelson and Zeckhauser 1988). |
| Net Benefits | By considering the costs (real or perceived) of switching from the status quo to a new position, decision-makers may choose not to make the switch (Polites and Karahanna 2012). Before making a switch to a new alternative, a decision maker goes through a rational decision-making process whereby the consideration of the difference between relative costs and benefits of change takes place (Kim and Kankanhalli 2009). |
| Transition Costs | Transition costs are incurred during the adaptation to the new situation and may enhance SQB if switching costs exceed the efficiency gain associated with a new alternative (Samuelson and Zeckhauser 1988; Kim and Kankanhalli 2009). |
| Uncertainty Costs | Uncertainty costs represent the psychological uncertainty or perception of risk associated with switching to a new situation (Samuelson and Zeckhauser 1988; Kim and Kankanhalli 2009). In addition, the individual's lack of information and expertise about the alternatives may impose search and analysis costs and lead to SQB (Lee and Joshi 2017). |
| Sunk Cost | Desire to justify previous commitments to a course of action by making a subsequent commitment (Samuelson and Zeckhauser 1988). |
| Regret Avoidance | Individuals are likely to avoid consequences in which they could appear to have made the wrong choice, even if the decision appeared correct in advance, given the information available at the time (Samuelson and Zeckhauser 1988). |
| Feel in Control | *"People resist change if they expect it to threaten the status quo, such as a potential loss of power or control over strategic organisational resources"* (Hsieh 2015, p. 517). |
| Cognitive Consistency | An individual finds it difficult to maintain two conflicting stances or ideas simultaneously and consequently seeks cognitive consistency, creating psychological commitment to the status quo (Samuelson and Zeckhauser 1988). |

*Table 1. Definitions of Status Quo Bias Theory Constructs*

## 2.1 Psychological Phenomena Promoting the Status Quo

"*The status quo refers to the existing state of affairs, or the way things are*" (Eidelman and Crandall 2009, p. 85). People associate long existence and persistence with goodness, righteousness, necessity, desirability, and legitimacy (Eidelman and Crandall 2014 p. 72, 93). According to Moshinsky and Bar-Hillel (2010) (as cited in Eidelman and Crandall 2014), "*because the status quo operates as a reference point from which change is considered, the costs of change should carry more weight than potential benefits, creating a relative advantage for the existing state of affairs*" (p. 59-60). "*Relative to alternatives, the status quo requires less effort, intention, control, and/or awareness for support and/or endorsement. As such, status quo maintenance is more ubiquitous and subtle than often believed*" (Eidelman and Crandall 2009, p.85). Though the long-existing, well-established institutions, rules, customs, and habits may not be ideal, changing them could be costly in terms of time, money, and effort (Eidelman and Crandall 2014; Niranga and Sedera 2022). A study by Parker (2021) reveals that a person exhibits status quo bias when he/she displays a fear of change and an emotional bias for choosing the known option. In life, "*maintaining the status quo is an attractive option as it does not require changing practice nor giving much thought to a decision*" (Parker 2021, p. 350). The study by Venkatesh et al. (2012) concluded that "*habit has both direct and mediated effects on technology use*" (p. 174). Table 2 describes the psychological phenomena favouring the status quo over alternatives.

| Psychological Phenomena | Description |
|---|---|
| Existence Bias | "*The tendency to assume that existing states of the world are good and right, to treat the ways things are as the way things ought to be*" (Eidelman and Crandall 2014, p. 55). |

| Longevity Bias | *"The tendency to assume that longstanding states of the world are good and right; longer existence is better"* (Eidelman and Crandall 2014, p. 55). |
|---|---|
| Primacy Effects | *"Information that comes early has an advantage over subsequent information"* (Eidelman and Crandall 2009, p.86) and *"Research has shown that early experiences are often remembered better than later experiences"* (Eidelman and Crandall 2009, p.86). |
| Anchoring | *"Because status quo alternatives come first, they are also likely to serve as a start value from which people may (or may not) move"* (Eidelman and Crandall 2009, p. 86). |
| Feature-Positive Effect | *"The status quo is more available and accessible than are other alternatives"* (Eidelman and Crandall 2009, p. 89). |
| Counterfactual Thinking | According to Eidelman and Crandall (2009), counterfactual thinking is a reasoning process in which an individual imagines an alternative to past or present factual events or circumstances. The ability to support a non–status quo stance requires an individual to invest some time in counterfactual thinking. A person is unlikely to favour a novel process, procedure, or politics without imagining what some of the outcomes would be. |
| Mere Exposure Effect | *"Since existing states will be encountered more frequently, mere exposure will lead them to be evaluated more positively"* (Eidelman and Crandall 2009, p. 92). |
| Post-decision Dissonance Effects | Preferring the status quo reduces the post-decision psychological tension (dissonance) that results from having to decide on a change (Eidelman and Crandall 2009). |
| A Mere Existence Bias | People consider the mere existence of something as evidence of its goodness, and an existing state is evaluated more favourably than an alternative (Eidelman and Crandall 2009). |

*Table 2. Psychological Phenomena Promoting the Status Quo*

# 3 Methodology - Systematic Literature Review

Our systematic literature review aims to develop a clear understanding of the current body of knowledge on the phenomenon of the status quo bias in users' information systems adoption and continuance intentions. Webster and Watson (2002) recommend a systematic literature review to gain a comprehensive overview of existing research on a given topic. Our review can be classified as a qualitative systematic review. In such a review, the reviewers *"might use various content analysis methods such as groupings, clusters, frameworks, classification schemes, and tabulations of characteristics to summarise the findings of the selected studies, narratively integrate the cumulative evidence, and arrive at conclusions and/or recommendations"* (Paré et al. 2015, p. 188). Status quo bias of users when adopting and using information systems presents an emerging issue (Chi et al. 2020; Oschinsky et al. 2021; Sedera et al. 2021) that would highly benefit from knowledge accumulation based on the integration of previous studies and findings (Paré et al. 2015) to map and assess the research area. Following the recommended guidelines on conducting systematic literature reviews (Okoli 2015; Paré et al. 2015; Webster and Watson 2002), this review covers relevant publications in peer-reviewed journals, conference proceedings and book chapters.

## 3.1 Search Process

The review topic, the status quo bias in users' information systems adoption and continuance intentions, can be considered an interdisciplinary research area drawing knowledge from disciplines such as management, IS, organisational behaviour and psychology. Therefore, we decided to use the databases SpringerLink, ProQuest Central, ScienceDirect (Elsevier), Emerald Insight, AIS eLibrary, Taylor & Francis Online, Web of Science Core Collection, Business Source Ultimate, ACM digital library, Cambridge core, JSTOR, SAGE journals and Oxford journals. In the first step, in order to perform an exhaustive search, we used the keyword "status quo bias". On all the databases, we searched the title, abstract and keywords. In the first step, we aimed to obtain a list of all the published research related to status quo bias.

The search on AIS eLibrary for the keyword "status quo bias" resulted in 75 papers. We then reviewed each paper based on screening the abstract. This resulted in the selection of 13 papers. At this initial stage, for a paper to be selected, it must be in English and require a connection to status quo bias and information systems.

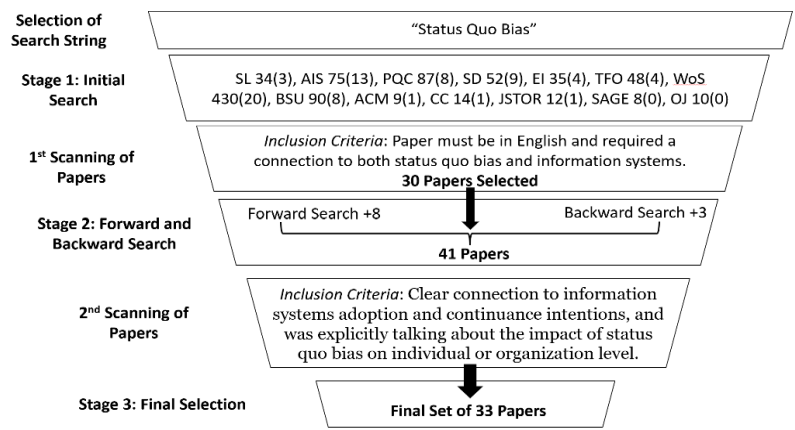

*Figure 1. Overview of Literature Review Process*

We then searched the databases SpringerLink (SL) (34 hits and 3 selected papers), ProQuest Central (PQC) 87(8), ScienceDirect (Elsevier) (SD) 52(9), Emerald Insight (EI) 35(4), Taylor & Francis Online (TFO) 48(4), Web of Science Core Collection (WoS) 430(20), Business Source Ultimate (BSU) 90(8), ACM digital library 9(1), Cambridge core (CC) 14(1), JSTOR 12(1), SAGE journals 8(0) and for the Oxford journals (OJ) 10(0). After removing duplicates (42), the initial stage resulted in 30 papers based on 904 initial hits. A forward (using Web of Science) and backward search of the 30 selected papers added eleven more articles, resulting in a set of 41 papers at the end of stage two of the search process. From the 41 papers, we moved on to the third screening stage, in which the entire paper was read. Regarding the inclusion criteria, we checked if the paper had a clear connection to information systems adoption and continuance intentions and explicitly discussed the impact of status quo bias on the individual or organisation level. Regarding our exclusion criteria, we removed all those papers that focused only on the "status quo" and technologies that are not related to information systems. We also checked if the paper was published in a peer-reviewed journal, conference, or book for the corresponding area. Completing the final stage resulted in 33 papers being retained as the final set. An overview of the process is shown in Figure 1.

## 3.2  Analysis

We performed a comprehensive literature review using the selected papers to extract information such as publication details, theories used, research questions and objectives, methodologies used, future directions, results, factors reducing user resistance to IS implementation, and use of status quo bias theory constructs. To analyse the final set of 33 papers, we created Excel tables to extract critical information about the papers. In the first table, we documented the author details, publication title and the publication outlet (name of the journal, conference, or the book publisher). In the second table, we included the theory/model/framework used in the research, the research question/objective of the research, and the methodology used. We used the third table to describe the application domain (health information systems, enterprise resource management systems, etc.), results, and future research agenda. In the fourth table, we documented the main contributions of the research. Intending to clarify the current body of knowledge on the concept of status quo bias, we started to work with the fundamental building blocks of Status Quo Bias Theory (Samuelson and Zeckhauser 1988) and added them as columns to our fifth analysis table. The fifth table is a concept matrix in which we documented whether the researchers have studied the Status Quo Bias Theory constructs (Loss Aversion, Net Benefits, Transition Costs, Uncertainty Costs, Sunk Cost, Regret Avoidance, Feel in Control, and Cognitive Consistency) (Samuelson and Zeckhauser 1988) and discovered any additional constructs. In the sixth table, we documented the factors reducing user resistance to IS implementation and their descriptions.

Paré et al. (2015) state that "*the defining element of qualitative systematic reviews is the adoption of a textual approach in the process of analysis and synthesis*" (p. 188.) Therefore, in order to answer the RQ1, we synthesised information based on the questions: Apart from SQBT, what are the theories, models and frameworks used to study SQB in IS?; What are the most common research methodologies applied to study SQB in IS?; What are the most significant results obtained in SQB in IS studies?; What are the levels of SQB influence?; What factors reduce user resistance to information systems implementation?; and How is SQBT applied in SQB in IS research? In order to answer the RQ2, we synthesised information based on the questions: What are the new SQBT constructs proposed in publications? And how does the proposed framework guide future studies of SQB in IS adoption and continuance?

In the last step of the study, using the steps followed by Haskamp et al. (2021), to develop the 'Research Framework for Inertia in Digital Transformation', we logically combined SQBT Constructs, levels of SQB influence, psychological reasons behind SQB, the impact of SQB, and factors reducing user resistance to IS implementation to propose 'SQB in IS adoption and continuance Framework'.

Regarding the process of the analysis and synthesis, we were informed by earlier review endeavours conducted by Haskamp et al. (2021), Piccoli and Ives (2005), and Paré et al. (2007).

## 4   Results

The profile of selected papers shows that SQB in IS is a growing stream of research that is gaining momentum and attention from the IS research community. In particular, the last few years have seen a substantial increase in publication volume (21 out of 33 papers were published after 2015), with papers published in top-tier IS conferences and journals such as the Association of Information Systems (AIS) Senior Scholars' Basket of Eight Journals (5 papers) and AIS premier information systems conferences (ICIS, ECIS and PACIS) (7 papers). Our findings are organised along with the ideas that stemmed from our literature review and that emphasise (1) levels of SQB influence, (2) new SQBT constructs proposed in publications, and (3) factors reducing the user resistance to information systems implementation.

### 4.1   The levels of SQB influence

According to Salahshour et al. (2018), research conducted on technology adoption "*aimed to understand, predict, and explain variables influencing adoption behaviour at the individual as well as organisational levels to accept and use technological innovations*" (p. 361). We summarise the studies into two categories: individual level and organisation level.

#### 4.1.1   Individual Level

According to SQBT, individuals are biased toward maintaining the status quo, that is, toward "*doing nothing or maintaining one's current or previous decision*" (Samuelson and Zeckhauser 1988, p. 7). "*The SQB perspective assumes that individual decision-makers are biased toward maintaining the status quo*" (Chi et al. 2020, p. 2). Individuals' reluctance to adopt new technology can arise because they are skeptical of these changes, worried about losing individual autonomy, and fear being replaced if they do not master the new technology (Oschinsky et al. 2021). "*At an individual level, change of technologies is not always welcome*" (Mahmud et al. 2017, p. 165). Individuals prefer the status quo to avoid change and an unknown outcome unless the perceived benefits clearly outweigh the perceived disadvantages (Heidt et al. 2017).

Individual-level attitude toward a change from the status quo will be positive if the change is perceived to offer relative advantage or value over and above the present situation (Kim and Kankanhalli 2009). According to Bhattacherjee and Hikmet (2007) and Recker (2014), accepting innovation and maintaining existing habits are the two factors that influence an individual user's decision to adopt new IT products. Individual-level resistance intentions have important ramifications for management decisions and organisational change (Recker 2014). "*The SQB perspective provides a context-dependent set of theoretical explanations for why an individual may remain in a status quo state even in the presence of better alternatives*" (Polites and Karahanna 2012, p. 22). Due to perceived costs of transitioning, uncertainty about the benefits of alternative systems, incorrectly factoring in sunk costs or a desire to maintain cognitive consistency, individual decision-makers may be biased toward maintaining the status quo (Polites and Karahanna 2012).

#### 4.1.2   Organisational Level

Implementing information technologies involves overall change management on an organisational level (Nugawela and Sedera 2021; Oschinsky et al. 2021) to adapt to substantial changes in technology, tasks, and people (Kim and Kankanhalli 2009). To implement a change to an existing IS, organisations need to change existing processes and structures. Strong organisational persistence of existing function (status quo) is essential to understanding the resistance to IS use within organisations (Polites and Karahanna 2012). In an organisation, "*habitual use of an existing IS can be a major source of inertia when a new system is* introduced" (Polites and Karahanna 2012, p. 26). The research conducted by Fan et al. (2015) regarding cloud system adoption discovered that perceived risk is positively related to organisational resistance, and perceived value is negatively related to organisational resistance.

## 4.2 New SQBT constructs proposed in publications

### 4.2.1 Work Impediments

According to the studies conducted by Recker (2014) and Sedera et al. (2021), work Impediments cause status quo bias. Recker (2014) defines work impediment as "*the individual assessment of system use, in terms of a detriment to work task performance due to a need to comply with the ineffectual requirements of the system use*" (p. 6). The relationship between SQB in IS and work impediments are established by analysing data collected from agri-business CEOs by Sedera et al. (2021). "*Work impediments in agriculture due to the use of IT can be defined as a detriment to an employee's daily job-related task performance resulting from having to use IT for farm activities*" (Sedera et al. 2021, p. 6). "*Perceived work impediment may exert a different, perhaps stronger influence on discontinuance intentions*" (Recker 2014, p. 9).

### 4.2.2 Domain Uncertainty

According to the study of agri-business CEOs conducted by Sedera et al. (2021), Domain Uncertainty cause status quo bias when using and adopting information technology in agricultural organisations. In agricultural organisations, uncertainties caused due to production, market, institutional, personal, financial, technological, and asset factors create domain uncertainty which raises SQB towards IT adoption (Sedera et al. 2021).

### 4.2.3 Anchoring effects

In their research about using the status quo bias perspective in IS research, Lee and Joshi (2017) have identified that anchoring effects create SQB in IS. "*Anchoring effects refer to individuals' propensity of setting a starting value and then assessing changes with reference to the initial state*" (Lee and Joshi 2017, p. 740). Furthermore, since implementing a new IS usually involves the replacement of a currently used IS, "*users may form expectations for the new IS based on their knowledge and experiences of past implementations*" (Lee and Joshi 2017, p. 743). However, according to the literature review conducted by Lee and Joshi (2017), prior IS adoption and resistance studies have not investigated anchoring effects.

## 4.3 Factors reducing the user resistance to IS implementation

Our analysis revealed eight factors that reduce the user resistance to information systems implementation (Table 3) that we introduce. In addition, we provide a summary of studies that support factors that reduce user resistance to IS (Table 4).

| Factor | Description |
|---|---|
| Switching benefits | Switching benefits indicate value associated with switching from the status quo to the new IS (e.g., improved quality of work, performing tasks more quickly) (Kim and Kankanhalli 2009). |
| Perceived value | Perceived value is the perceived benefits proportional to the costs of new IS-related change (Kim and Kankanhalli 2009). |
| Perceived usefulness | Perceived usefulness is "*the degree to which a person believes that using a particular system would enhance his or her job performance*" (Davis 1989, p. 320). |
| Favourable colleague opinion | Kim and Kankanhalli (2009) define the favourable colleague opinion as "*the perception that colleagues favour the changes related to a new IS implementation*" (p. 573). The research conducted by Eckhardt et al. (2009) shows significant impacts of social influence from workplace referents on IT adoption. |
| Self-efficacy for change | Bandura (1995) (as cited in Kim and Kankanhalli 2009) defines self-efficacy for change "*as an individual's confidence in his or her own ability to adapt to the new situation*" (p. 573). |
| Organisational support for change | Organisational support for change is defined as "*the perceived facilitation provided by the organisation to make users' adaptation to new IS-related change easier*" (Kim and Kankanhalli 2009, p. 573). |
| Value for others | Value for others "*refers to the estimated benefit for external actors*" (Oschinsky et al. 2021, p. 4). |
| Perceived ease of use | Davis (1989) (as cited in Bhattacherjee and Hikmet 2007) defines perceived ease of use as "*the extent to which users believe that their system usage will be relatively free of effort*" (p. 728). |

*Table 3. Description of factors reducing the user resistance to information systems implementation*

| Factor | Supported Studies |
|---|---|
| Switching benefits | Chang (2020); Heidt et al. (2017); Kim and Kankanhalli (2009); Mahmud et al. (2017); Oschinsky et al. (2021) |
| Perceived value | Balakrishnan et al. (2021); Chi et al. (2020); Cho et al. (2021); Fan et al. (2015); Gong et al. (2010); Hsieh et al. (2014); Hsieh (2015); Hsieh and Lin (2018); Hsieh and Lin (2020); Kim and Kankanhalli (2009); Mahmud et al. (2017); Oschinsky et al. (2021); Shankar and Nigam (2021); Shirish and Batuekueno (2021) |
| Perceived usefulness | Balakrishnan et al. (2021); Bhattacherjee and Hikmet (2007); Chi et al. (2020); Cho et al. (2021); Hsieh et al. (2014); Hong et al. (2011); Polites and Karahanna (2012); Recker (2014); Tsai et al. (2019) |
| Favourable colleague opinion | Cho et al. (2021); Kim and Kankanhalli (2009); Oschinsky et al. (2021); Prakash & Das (2021); Shirish and Batuekueno (2021) |
| Self-efficacy for change | Bhattacherjee and Hikmet (2007); Cho et al. (2021); Hsieh (2015); Kim and Kankanhalli (2009); Shirish and Batuekueno (2021) |
| Organisational support for change | Cho et al. (2021); Kim and Kankanhalli (2009); Oschinsky et al. (2021); Shirish and Batuekueno (2021) |
| Value for others | Oschinsky et al. (2021) |
| Perceived ease of use | Balakrishnan et al. (2021); Bhattacherjee and Hikmet (2007); Chi et al. (2020); Cho et al. (2021); Hong et al. (2011); Recker (2014); Tsai et al. (2019) |

*Table 4. Supported studies for factors reducing the user resistance to information systems implementation*

### 4.4 SQB on IS adoption and continuance Framework

Based on the findings from the systematic literature review and SQBT, we propose a framework to investigate the SQB in IS adoption and continuance in organisations. The framework aims to provide an overview of the psychological phenomena that promote SQB, SQBT constructs, the effect of SQB on the individual and organisational level, factors reducing the user resistance to IS implementation, and how the aforementioned factors influence the users' IS adoption and continuance intentions. In addition, we introduce additional SQB constructs that are proposed in reviewed studies. Our framework is multidimensional since it describes the antecedents of SQB and the factors that will counter the impact of SQB on users' IS adoption and continuance intentions. We do not claim our framework to be a comprehensive overview regarding SQB in users' IS adoption and continuance intentions. We instead see it as a starting point for the researchers to add more dimensions, constructs, and relationships in future studies.

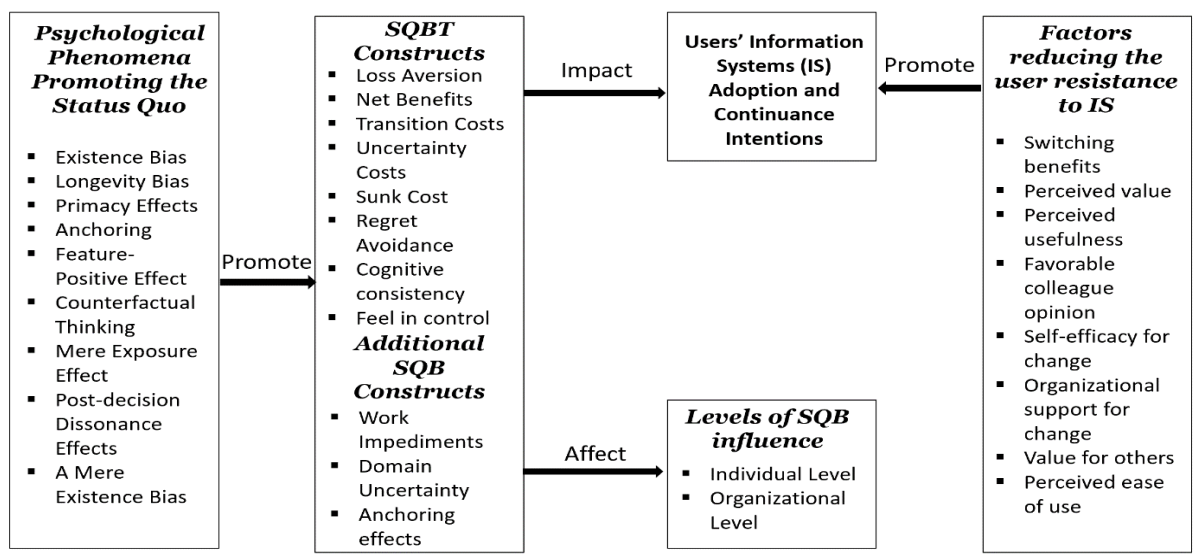

*Figure 2. SQB in IS adoption and continuance Framework*

## 5  Conclusion

In this review, we sought to understand emerging research on SQB in users' IS adoption and continuance intentions. In our first research question, our review intended to investigate the current body of knowledge on the role of SQB in the context of IS adoption and continuance. Based on our literature review, we present a framework for research on SQB in IS, including 'Psychological Phenomena Promoting the Status Quo', 'SQBT constructs and new SQB constructs', 'levels of SQB influence', and 'Factors reducing the user resistance to IS implementation', answering the second research question. While SQB is mentioned as a core barrier to IS implementation in highly cited publications by Kim and Kankanhalli (2009) and Polites and Karahanna (2012), current research has only started to better understand the phenomena in the context of specific domains of IS implementations (Agribusiness - Sedera et al. (2021), Health record systems - Cho et al. (2021), Enterprise resource planning - Alzahrani et al. (2021), Chang (2020) and Mahmud et al. (2017), Decision support systems - Prakash & Das (2021), Health information technology - Hsieh and Lin (2020), Chi et al. (2020), Hsieh and Lin (2018), and Tsai et al. (2019), IS in Government institutions - Oschinsky et al. (2021) and Schirrmacher et al. (2019), Artificial Intelligence based IS - Balakrishnan et al. (2021)). The detailed analysis of the literature identifies the role of SQBT in users' IS adoption and continuance intentions.

### 5.1  Research and Practical Implications

We see our proposed framework as a starting point to enhance the understanding of SQB in the context of users' IS adoption and continuance intentions. First, by presenting this framework, we expect to report to future IS scholars to acknowledge the role of SQB in studies related to users' IS adoption and continuance intentions. Second, through the literature review, we documented three new SQB constructs (Lee and Joshi (2017); Recker (2014); Sedera et al. (2021)) that could extend the theoretical foundations of SQBT.

First, this study provides guidance for practitioners to better understand the effects of SQB on user resistance to IS implementations. Second, understanding the 'Psychological Phenomena Promoting the Status Quo' can help devise intervention mechanisms to minimise the effects of status quo bias. Third, the study insights about 'Factors reducing the user resistance to IS implementation' would provide guidance for practitioners to promote users' IS adoption and continuance intentions within the organisations.

### 5.2  Limitations and Future Work

This study has several limitations. First, while we are aware of the broad range of research on SQB ranging from psychology, and management to IS, we followed a relatively narrow conceptualisation of users' IS adoption and continuance intentions which might have led to the exclusion of articles that may be relevant. Second, the notion of SQB is discussed in certain publications under the term 'inertia', which may present research opportunities that could be incorporated into our framework. Third, given the available space, our paper lacks an extensive elaboration on the findings of the literature analysis. Finally, our proposed framework intends to foster initial comprehension of 'SQB in users' IS adoption and continuance intentions' rather than an all-encompassing and potentially more complex explanation.

Future studies could include empirical research on generalising the validity of three new SQB constructs. Second, the study conducted by Hoehle et al. (2012) with internet banking users in New Zealand has identified that continuous trust is a significant contributor to ongoing intention to use the system. A future study could add specific findings on the impact of 'continuous trust' on SQB in IS. Third, future studies could empirically validate the relationships between SQBT constructs and the 'Psychological Phenomena Promoting the Status Quo'. Fourth, more research is needed on the effects of SQBT constructs on organisational level SQB. Finally, researchers could validate the moderating effect of the 'Factors reducing the user resistance to IS implementation' on SQB.